\documentclass[useAMS,usenatbib]{mn2e}
\usepackage{lscape}
\usepackage{wrapfig}
\usepackage{deluxetable}
\newcommand\xmm{{\it XMM-Newton}}

\newcommand\kms{\ifmmode {\rm~km\ s}^{-1} \else ~km s$^{-1}$\fi}
\newcommand\Hunit{\ifmmode {\rm~km\ s}^{-1}\ {\rm Mpc}^{-1}
        \else ~km s$^{-1}$ Mpc$^{-1}$\fi}
\newcommand\ctssec{\ifmmode {\rm~count\ s}^{-1} \else ~count s$^{-1}$\fi}
\newcommand\ergsec{\ifmmode {\rm~erg\ s}^{-1} \else
        ~erg s$^{-1}$\fi}
\newcommand\funit{\ifmmode {\rm~erg\ s}^{-1}\;{\rm cm}^{-2} \else
        ~ergs s$^{-1}$ cm$^{-2}$\fi}
\newcommand\phflux{\ifmmode {\rm~photon\ s}^{-1}\;{\rm cm}^{-2}
        \else   ~photon s$^{-1}$ cm$^{-2}$\fi}
\newcommand\efluxA{\ifmmode {\rm~erg\ s}^{-1}\;{\rm cm}^{-2}\;{\rm
        \AA}^{-1} \else ~erg s$^{-1}$ cm$^{-2}$ \AA$^{-1}$\fi}
\newcommand\efluxHz{\ifmmode {\rm~erg\ s}^{-1}\;{\rm cm}^{-2}\;{\rm
        Hz}^{-1} \else ~erg s$^{-1}$ cm$^{-2}$ Hz$^{-1}$\fi}
\newcommand\cc{\ifmmode {\rm~cm}^{-3} \else cm$^{-3}$\fi}
\newcommand\FWHM{\ifmmode {\rm~FWHM} \else ${\rm~FWHM}$\fi}
\newcommand\Msun{\ifmmode M_{\odot} \else $M_{\odot}$\fi}
\newcommand\Lsun{\ifmmode L_{\odot} \else $L_{\odot}$\fi}

\newcommand\hbeta{\ifmmode {\rm H}\beta \else H$\beta$\fi}
\newcommand\Kalpha{\ifmmode {\rm K}\alpha \else K$\alpha$\fi}
\newcommand\nh{\ifmmode N_{\rm H} \else N$_{\rm H}$\fi}
\usepackage{graphicx}
\usepackage{color}

\title{Mid-Infrared and X-ray luminosity correlations of X-ray point sources in NGC 1399}

\author[P. Shalima et al.]{P. Shalima $^1$\thanks{E-mail: shalima.p@gmail.com (PS)}, V. Jithesh$^2$, K. Jeena$^4$, R. Misra$^{1}$, S. Ravindranath$^1$, 
\newauthor
 G. C. Dewangan$^{1}$, C. D. Ravikumar$^2$ and B. R. S. Babu$^{3}$\\
$^{1}$ Inter-University Centre for Astronomy and Astrophysics, Post Bag4, Ganeshkhind, Pune-411007, India\\
$^{2}$ Department of Physics, University of Calicut, Malappuram-673635, India\\
$^{3}$ Department of Physics, Sultan Qaboos University, Muscat, Oman\\
$^{4}$ Department of Physics, Providence Womens' College, Malaparamba, Calicut-673009, India}

\begin{document}
\maketitle
\label{firstpage}

\begin{abstract}
It is known that the IR and X-ray luminosities of Active Galactic Nuclei (AGN) are
correlated with $L_{IR} \sim L_{X}$. Moreover, the IR flux ratio between the
$5.8$ and $3.6\mu$m bands is a good distinguishing characteristic of AGN or AGN-like behaviour.
On the other hand, Galactic X-ray binaries (GXB) are under-luminous in the IR with
$L_{IR} << L_X$. Since Ultra-luminous X-ray sources in nearby galaxies may be an intermediate
class  between AGN and GXB, it is interesting to study if their IR properties indicate which
kind of objects they resemble. We use Spitzer IRAC images to identify mid-IR counterparts
of bright X-ray sources, detected by {\it Chandra} in the elliptical galaxy NGC 1399.
We find that for sources with AGN-like IR flux ratios, the IR luminosity strongly correlates
with that in X-rays, $L_{IR} \sim L_X$, while for the others, there is no correlation between the
two. Some of the former objects may be background AGN. If they are not strongly contaminated
by background AGN, this result extends the IR-X-ray luminosity correlation down to $L_X \sim 10^{39}$
ergs/s. We calculate their $g-z$ colours and find that the bright X-ray sources with IR
counterparts are typically blue in optical color. This is in contrast to typical X-ray
sources, without IR counterparts which have predominantly red optical counterparts. We highlight the need for IR or
optical spectra of these sources to distinguish background AGN and unveil the effect of the
X-ray emission on the different environments of these systems.
\end{abstract}


\section{Introduction}

{\it Chandra}  and \xmm{} have detected several off-nuclear X-ray point sources in nearby  galaxies. 
Many of these sources are expected to be like typical X-ray binaries found in our Galaxy. However, unlike
typical Galactic X-ray binaries (GXB), some of them have luminosities in excess of $10^{39}$ ergs s$^{-1}$ and are called Ultra-Luminous X-ray sources (ULXs). As these ULXs 
have a luminosity higher than the Eddington 
limit for an accreting ten solar mass black hole, they may be harbouring black 
holes in the mass range of 100 - 10 $^{5} \Msun $ which are intermediate between 
the stellar mass black holes of Galactic X-ray binaries (GXB) and the 
super-massive black holes of Active Galactic Nuclei (AGN) \citep{Col99,Mak00}. 
 Spectral analysis of some ULXs seem to favour an IMBH 
interpretation, though not conclusively \citep{Miller03,Miller04,Dev08}. 
On 
the other hand, they could also be mildly beamed, super-Eddington accretion 
disks around stellar mass black holes as supported by recent spectral analysis 
\citep{Shak73, King08, Kuncic2007, Feng2011}.  
Similar to GXB, some ULXs have exhibited spectral state transitions \citep{Col99,Kub01,Fen06} associated with changes in temporal behaviour 
\citep{Dew10}. However, unlike GXB, the spectra of many ULXs exhibit a high 
energy cutoff at $\sim 6$ keV \citep{Agr06,Dew06a,Sto06}. Again, like GXB, 
 a handful of ULXs exhibit quasi-periodic oscillations e.g., 
M~82~X-1 \citep{Str03}, Holmberg IX~X-1 \citep{Dew06b}, NGC~5408~X-1 
\citep{Str2007}, NGC~6946~X-1 \citep{Rao2010} and X42.3+59 
\citep{Feng2010} at lower frequencies ($\sim 100$ mHz), which  indicates that they may indeed harbour IMBH. 
Thus, while the X-ray luminosity, spectral and temporal properties of a few 
ULXs suggest that they are intermediate objects between GXB and AGN, the definitive measurement of their black hole masses is yet to be made and hence their nature still remains largely mysterious.  Moreover, since these X-ray sources
are in general variable, some of the lower luminosity sources may be ULXs that have not been observed during their
 high flux period. It is important 
therefore, to study these sources at other wavelengths which may provide some 
distinguishing characteristics that can identify them to be more like AGN or 
GXB. 

To understand 
the nature of the environment in which these X-ray sources are created,  
there  have been extensive optical studies on the properties of the globular clusters that 
host an X-ray source \citep{Kim06,Kim09}.
There have  also been several studies and detections of optical counterparts of ULXs \citep[e.g.][]
{Liu04,Kuntz05,Ramsey06,Terashima06}. While a few have been identified as O-type stars 
\citep{Liu02,Liu07}, the majority are star clusters \citep[e.g.]{Goad02,Ptak06}. 
 Unless its companion is a bright O-type star, an object 
similar to a GXB will be relatively faint in optical to be detected. 
 For AGN including blazers, the ratio of the optical to X-ray flux can vary from 
0.1-50 \citep{Sto91}. However, based on  optical photometry alone (i.e. without detailed spectral 
information), it is difficult to discern and identify an AGN-like optically bright ULX, from a 
source in a globular cluster. Nevertheless, the absence of a bright optical counterpart for a 
ULX can be effectively used to impose a strong upper limit on the mass of the black hole,  $M_{\rm BH} < 1300 M_\odot$ \citep{Jith11}. 

 Another wavelength regime that could provide useful information regarding these sources is the infrared (IR).
For GXB, the IR emission is correlated with the X-ray and is believed to originate
from the companion star, the outer regions of the accretion disk and/or a jet \citep{Russell06}.  \citet{Harrison2011} have found periodic mid-IR emission in a LMXB GX17+2 which was consistent with synchrotron emission from the jet. For another LMXB GS 2023+338, \citet{Muno2006} found the mid-IR emission to be originating from dust in the accretion disc and follow the $\nu^{-2}$ law corresponding to the Rayleigh-Jeans tail of Black Body emission. However the
IR luminosity is substantially weaker than the X-ray luminosity i.e. $L_{IR} << L_{X}$, and hence 
it is not detectable in extra-galactic sources. 

 The mid-IR (3-40$\mu$m) emission in star-forming galaxies is also known to be correlated with X-ray emission but the mid-IR flux is 3-4 magnitudes higher than the X-ray and is dominated by PAH (Polycyclic Aromatic Hydrocarbon) emission \citep{Symeonidis}. This correlation exists even at small scales \citep{Tyler2004} and is an indication of recent active star-formation \citep{Kennicutt1998}.

 A strong correlation between the X-ray and 
IR luminosities, $L_{IR} \sim L_{x}$ \citep[e.g.][]{asmus11,krabbe,Lutz04} is also seen in AGN. Here the IR emission 
is due to reprocessing of the UV/X-ray emission from a dusty environment, except in jet-dominated blazars \citep{Matsuta}. The difference in the 
ratio of the IR to X-ray luminosities of XRB and AGN, reflects a difference in the environment 
of these two types of sources, which may otherwise be black hole mass scaled versions of each other.  
AGN have red and featureless spectra in the mid IR 
\citep[e.g][]{Houck05,Hao05,Weedman05} and they have unique colours in this wavelength region which 
can be effectively used to identify them \citep{Poll06,Lacy04,Stern05,Hat05}. For example, a positive 
flux ratio (in logarithmic units) between $5.8$ and $3.6 \mu$m bands implies that the IR emission 
is from an AGN \citep{Lacy04}. The $5.8 \mu$m channel contains the 6.2$\mu$m PAH feature \citep{Leger1984,Allamandola1989} with only a 20\% stellar contribution. On the other hand, the 3.6$\mu$m channel consists mainly ($>$ 90\%) of stellar emission \citep{Wang2004}. Thus, such flux ratios may be taken as an indication of a dusty environment 
around a bright UV/X-ray source.

The IR emission of  X-ray sources in nearby galaxies can provide valuable information regarding their
environment. While some of them are expected to be similar to Galactic X-ray binaries
where the IR is emitted from an outer disk or jet and having $L_{IR} << L_{X}$, there may be another
type of sources where, like AGN, the IR  may be due to reprocessed emission with $L_{IR} \sim L_{x}$.
  Moreover, since the IR-X-ray 
correlation in AGN, has been recently shown to extend to low luminosity AGN ($L_{X} \sim 10^{42}$ 
ergs/s), it is interesting to know whether there are sources at still lower X-ray luminosities 
($L_{X} \sim 10^{37-40}$ ergs/s) for which the correlation holds. 

Most of the IR studies of X-ray sources in nearby galaxies have been based on spectral studies. 
For 6 ULXs in NGC4485/4490, \citet{Vaz07} used Spitzer IRS spectra in order to derive  
IR spectral diagnostics that are characteristic of X-ray/UV heating of the surrounding gas 
and dust. They made use of the spectral resolution of the IRS instrument  to estimate line ratios 
such as [NeIII]/[NeII] and [SIII]/[SiII] for the ULXs. These ratios, as suggested by \citet{Dale06} 
are useful in separating accretion-powered from star-formation powered systems. \citet{Vaz07} found 
that 5 of the ULXs have these ratios similar to AGN while one of them may correspond to a star 
forming region. \citet{Vaz07} also modelled the continuum and derived dust temperatures for the sources, 
from which they suggested that for the two sources having hotter dust components, the [SiII] lines may 
be originating from the accreting source itself.  Again, using the IRS instrument, \citet{Bergh10a}  
detected [OIV] 25.89 $\mu$m emission from the ULX Holmberg II. Since this emission is associated with 
high excitation levels seen in AGN, it is an indication that the ULX is influencing its environment. 
By a detailed photo-ionization modelling of the IR lines, \citet{Bergh10b} concluded that the 
luminosity and morphology of the [O IV] 25.89 $\mu$m emission line is consistent with photoionization 
by the soft X-ray and far ultraviolet (FUV) radiation from the accretion disk and inconsistent with 
narrow beaming. They also argued that in order to produce the observed [OIV] flux, the source should 
have a bolometric unabsorbed luminosity in excess of 10$^{40}$ ergs s$^{-1}$.

While such detailed analysis can be undertaken if high resolution spectra are available,
significant results can also be obtained by photometric measurements of IR counterparts 
of X-ray sources in nearby galaxies. Since the IR photometric data can be obtained for a larger 
number of X-ray sources in a galaxy, one can study statistical properties such as whether the X-ray 
luminosity correlates with the IR or not. To identify IR counterparts of X-ray sources unambiguously, 
the host galaxy should have a smooth continuum IR profile which can be modelled and subtracted out, 
to reveal the true point-like IR sources. Thus, a nearby elliptical galaxy which hosts a large number 
of X-ray sources would be ideal for such a study.  
NGC1399 is a bright elliptical galaxy in the centre of the Fornax cluster. It is unique in having 
a large number of X-ray sources with a significant fraction of them being ULXs  \citep{Swartz04}. 
 \citet{Angelini} have studied the X-ray sources in this galaxy and found that about 70\% of them are associated
with globular clusters, which implies that the X-ray sources are formed preferentially in globular clusters.

In this paper, we have analysed archival data of NGC1399 from Spitzer IRAC and {\it Chandra}, in 
order to study the IR counterparts of the X-ray sources in the galaxy. We have used the IR 
photometeric flux ratios to identify AGN-like sources and to see if the IR flux is correlated with 
that of the X-rays as seen in AGN.

\section{X-ray and Infrared data Analysis}

\subsection{Chandra data}

We have analysed a 56 ksec {\it Chandra} observation (ID : 319) of NGC 1399. The data reduction and 
analysis were done using CIAO 4.2, and HEASOFT 6.9.0. Using the CIAO source detection tool {\it 
celldetect}, a total of 120 X-ray point sources were extracted from the level 2 event list with 
{\it signal-to-noise} ratio of 3. Of these 120 sources, 118 were within the field of view of the 
Spitzer IRAC observation and hence these sources were considered as the sample.

For some of the 118 point sources, the net X-ray count is too small ($N_C < 50$) for spectral
analysis. Moreover, some sources, typically near the galaxy centre, are in regions of excessive 
diffuse emission. Taking these aspects into consideration, we were able to extract reliable spectra 
for 35 sources (Tables 1,2,and 3) out of a total of 118 which were in the FOV. The spectral analysis was done using XSPEC version 12.6.0, and 
the data were fitted in the energy range of $0.3-8.0$ keV. The inferred luminosity of a source 
depends on the spectral model used and hence following \cite{Dev08}, we fitted each source independently with an 
absorbed powerlaw and an absorbed disk blackbody spectral model. The model which provided a
lower $\chi^2$ was considered as a better representative and the corresponding X-ray luminosity
was ascribed to that source. Two of these sources required an additional diffuse thermal component 
which was modelled using the XSPEC model, mekal. For the rest of the sources, the X-ray counts
are taken to be a proxy for the luminosity of the source. For example, \citet{Angelini} considered
the X-ray sources of this galaxy and estimated a conversion factor between the count rate and luminosity. This
assumes that the faint sources have roughly the same spectra and can be represented by a power-law absorbed by
Galactic absorption. 

\subsection{Spitzer IRAC}

The IRAC instrument on board Spitzer provides mid-IR images at four 
wavelengths i.e. 3.6, 4.5, 5.8 and 8$\mu$m. The IRAC Post Basic 
Calibrated Data (PBCD) mosaic files of NGC1399 were obtained from the Spitzer 
Heritage Archive. We have used the data corresponding to AOR 5529856 which was 
part of the Spitzer Infrared Nearby Galaxies Survey \citep{Kenn03}. 
 The exposure time was 840 seconds. 

We used the X-ray positions of 118 X-ray sources near NGC1399 in order to look 
for IR counterparts. Out of the four wavelengths, 3.6 and 5.8$\mu$m were used in this analysis as only they had good coverage of locations of X-ray point sources. 

We used the 3.6$\mu$m image which has the highest spatial resolution and extracted sources 
at the 3-sigma level using the Source Extractor package of \citet{Bert96}. For 
sources which would otherwise be undetectable against the bright image of the galaxy, 
an isophotal model of the elliptical galaxy was subtracted from the original image 
using the IRAF package before extracting point sources.

\begin{figure}
\centering
\includegraphics[width=6.45cm,height=5.3cm]{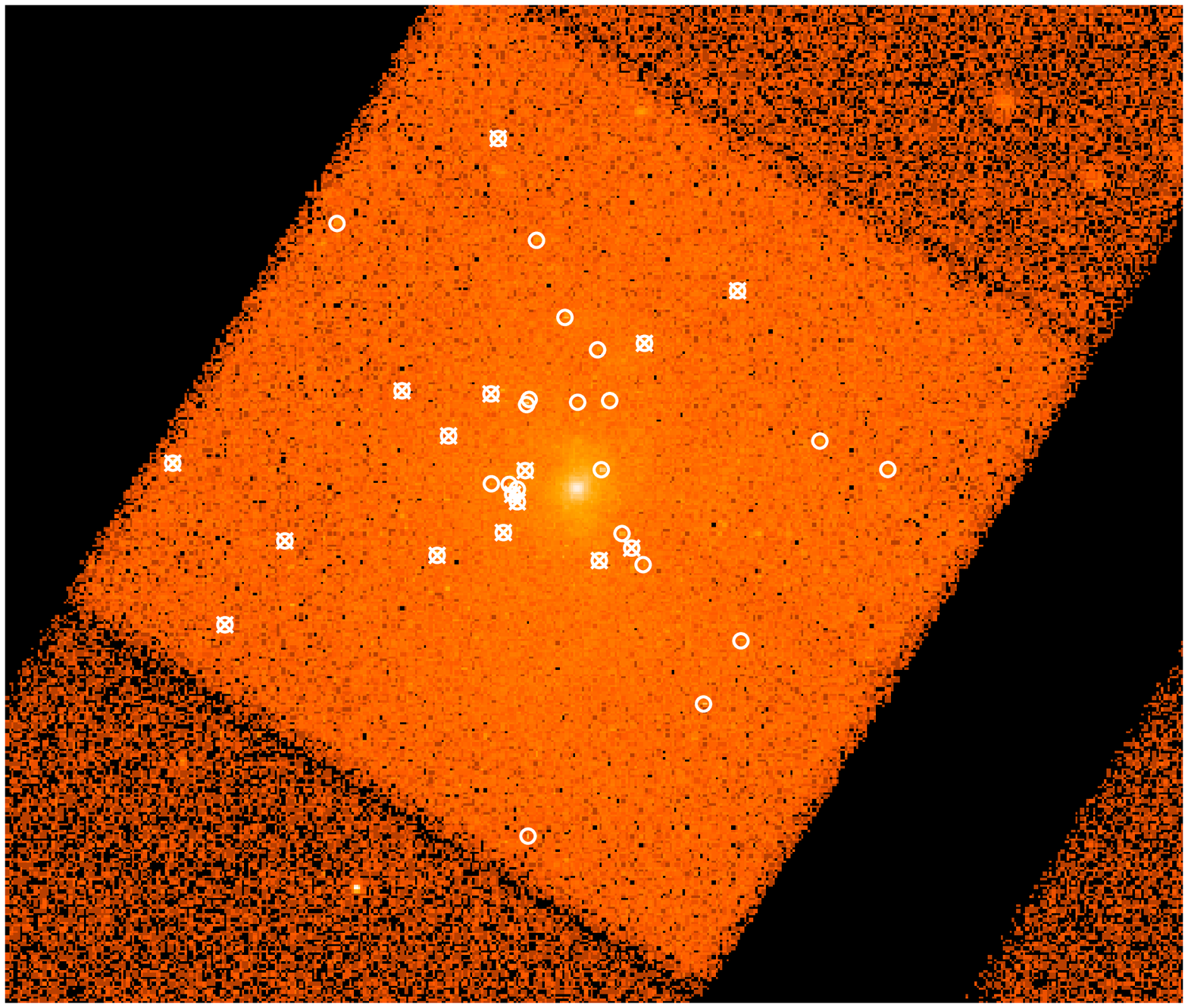}
\includegraphics[width=6.45cm,height=5.3cm]{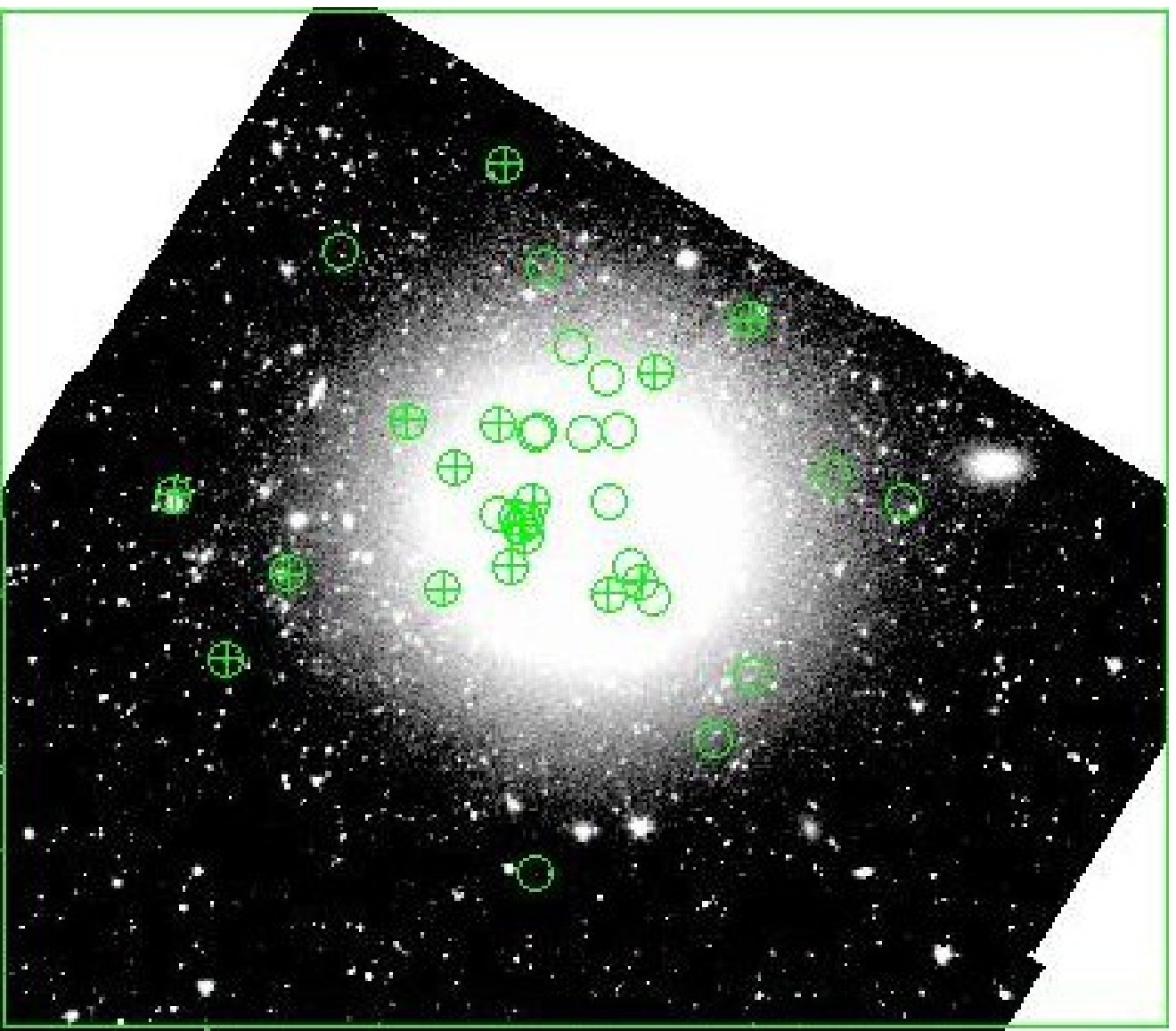}
\caption{{\it Chandra} (above) and {\it Spitzer} (below) image of NGC1399 with the X-ray sources with spectra (circles) and their IR counterparts (crosses) marked. The size of the images are 12 by 10 arcminutes}
\label{image1}
\end{figure}

 At the 3-sigma level, the source extractor detected 827 IR sources in the 3.6$\mu$m 
image of NGC 1399 over an area covering $12^{\arcmin}$ by $10^{\arcmin}$, which 
covers the coordinates of  118 X-ray  point sources (Figure \ref{image1}). 
 Thus the average number density of sources in the field is $\sim 1.9 \times 10^{-3}$ sources/arcsec$^2$.
The relative astrometry between {\it Chandra} and HST is around 1.5 arcsecs and the size of the
{\it Chandra} point spread function is around 0.5 arcsecs \citep[e.g.][]{Angelini}. Here, we assume that the relative astrometry between 
{\it Chandra} and {\it Spitzer} is around 2 arcsecs
and look for the IR counterparts of the X-ray sources. The  probability that an IR source
will fall within 2 arcsecs of an X-ray coordinate by chance is
roughly the IR source density times the area of a circle of 2 arcsec i.e. $\sim 2.4$ \%.
Of the 118 X-ray sources considered we may expect around 3 of them to have an IR source within 2 arcsecs just by chance.
Instead, we find that 33 of them have an IR counterpart. This implies that most of the IR counterparts are real and not chance coincidences.
 Figure \ref{image1}  show the Chandra and Spitzer images of the region with the X-ray sources and their IR counterparts marked.

 For all IR sources we used the 
APEX software available at the Spitzer website in order to derive the aperture fluxes as well as the uncertainties inside the extraction aperture. The flux uncertainties were obtained from error maps that were downloaded from the Spitzer website.

 For the aperture photometry, we used an aperture of radius 1.44$\arcsec$ and 1.49$\arcsec$, twice the FWHM of the 3.6 and 5.8$\mu$m channels respectively. For the X-ray sources that are not detected at 3.6$\mu$m,  
 we quote the 3-sigma upper limits at the locations. Luminosities ($\nu$f$_{\nu}$$\times$4$\pi$$D^{2}$) 
were then calculated using the distance to NGC 1399 to be $D = 19$ Mpc. The NASA/IPAC Extragalactic Database (NED)
shows that there are a large number of distance estimate for NGC 1399 ranging from $13.8$ to $24.9$ Mpc, and hence we have taken the average value of 19 Mpc. 

\begin{figure}
\centering
\includegraphics[width=9cm,height=9cm]{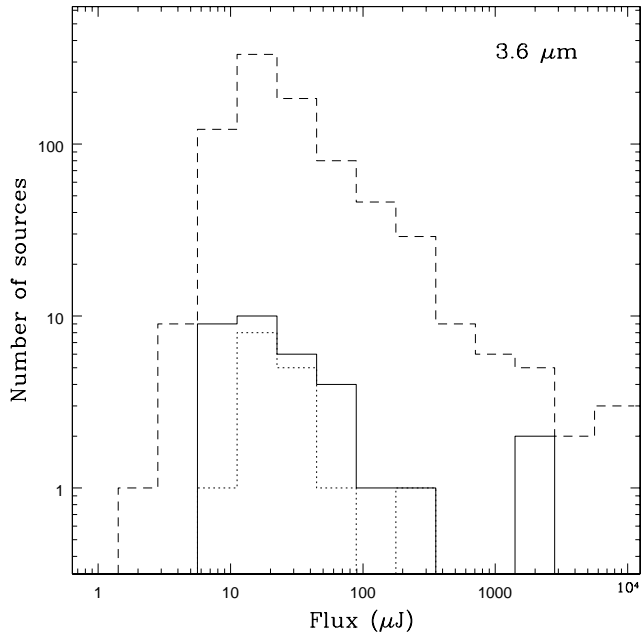}
\caption{Histogram of the IR luminosities of all sources detected in the IRAC image (dashed line) and the counterparts of X-ray point sources (solid line). The K-S test statistic is d=0.15 with a probability of 0.43. Sources with X-ray spectra are indicated by the dotted line. The K-S test statistic for these sources is d=0.13 (prob = 0.92).}
\label{hist1}
\end{figure}

\begin{figure}
\centering
\includegraphics[width=9cm,height=9cm]{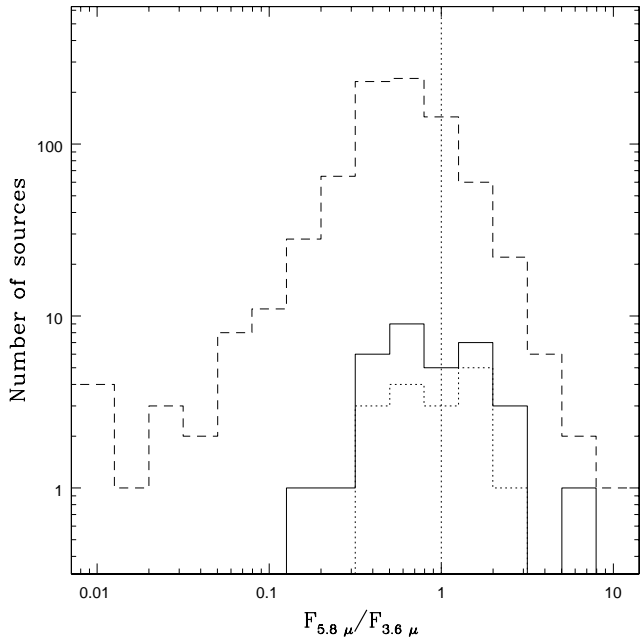}
\caption{Histogram of the IRAC colours of all sources detected in the IRAC image (dashed line) and the counterparts of X-ray point sources (solid line). Also plotted is the histogram for the 16 sources with X-ray spectra (N$_{C}>$50), which have detected mid-IR counterparts (dotted line). The vertical line demarcates the AGN-like from the non AGN-like sources.}
\label{histc}
\end{figure}

\section{Source Distributions}

In this section, we compare the distributions of all the IR and X-ray point sources in the
sample and compare them with those that have an IR/X-ray counterpart. This is useful to
see if the sources which are bright in both IR and X-rays are different from the
parent distribution. However, we note that neither the IR nor the X-ray source sample used
is complete. There are IR sources considered which are not in the {\it Chandra} field of view.
More importantly, detection of IR and X-ray point sources
depends on the level of the background  diffuse emission at that location
and for Chandra the off axis efficiency of point source detection is different than the on-axis one. 
Our interest here is to check whether on an average the properties of sources
 that are IR counterparts of X-ray sources, differ from the remaining sources identified in the IR images.

Figure \ref{hist1} 
shows the histogram of the IR 3.6 $\mu$m flux for all 827 sources in the field (dashed line) and that of the
33 IR sources detected in X-rays (solid line). A Kolmogorov-Smirnov (K-S) test, reveals that there
is no significant difference between the two distributions with K-S statistic $d = 0.15$ corresponding
to a probability of $0.43$. Throughout this work we consider a probability of 0.05 or less to be significant.
The dotted line in the figure shows the distribution of the 16 IR sources 
with X-ray counts $N_C > 50$, and that distribution is also not different from the parent one with
$d=0.13$ corresponding to a probability of $0.92$. Thus in terms of IR flux, there is nothing
special about the IR sources having bright X-ray emission.  

AGN have unique colours in the mid-IR range and in particular the
 IRAC flux ratio between the $5.8$ and $3.6 \mu$m bands, $F_{5.8}/F_{3.6} > 1$
is a good identifier \citep{Lacy04}. Figure \ref{histc} shows the histogram of this
ratio for the field 827 IR sources (dashed line) and for the 33 sources with X-ray emission (solid line).
The two distributions are significantly different with a K-S statistic $d = 0.29$ corresponding
to a probability of $0.006$ (assuming a level of significance of 0.05). This difference is primarily because the fraction of IR sources that are X-ray bright with 
$F_{5.8}/F_{3.6} > 1$ is $\sim 10$\% as compared to $\sim 5$\% for those 
that have $F_{5.8}/F_{3.6} < 1$. The dotted line represents the 16 sources with X-ray counts, $N_C > 50$
which indicates that they have similar IR flux ratio as sources with $N_C < 50$.

\begin{figure}
\centering
\includegraphics[width=9cm,height=9cm,angle=0]{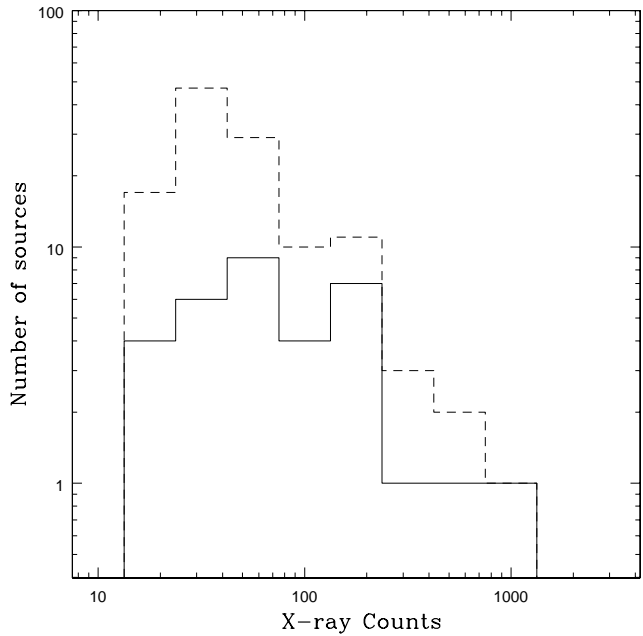}
\caption{Histogram of X-ray counts for the X-ray point sources (dashed Line) and their 
IR counterparts.}
\label{histxc}
\end{figure}

Next, we checked the X-ray intensity (or counts) of the X-ray sources that have IR emission.  
Figure \ref{histxc} shows the distribution of the 118 X-ray points sources (dashed line) compared 
with the 33 sources with IR emission. The K-S statistic is $d = 0.24$ with a probability of $0.089$. 

Thus, although the probability of an IR point source to be X-ray bright does not seem to depend on 
the IR flux, it increases with the IR flux ratio $F_{5.8}/F_{3.6}$. Moreover a brighter X-ray point
source is slightly more likely to have an IR counterpart than a fainter one.

\begin{figure}
\centering
\includegraphics[width=9cm,height=9cm,angle=0]{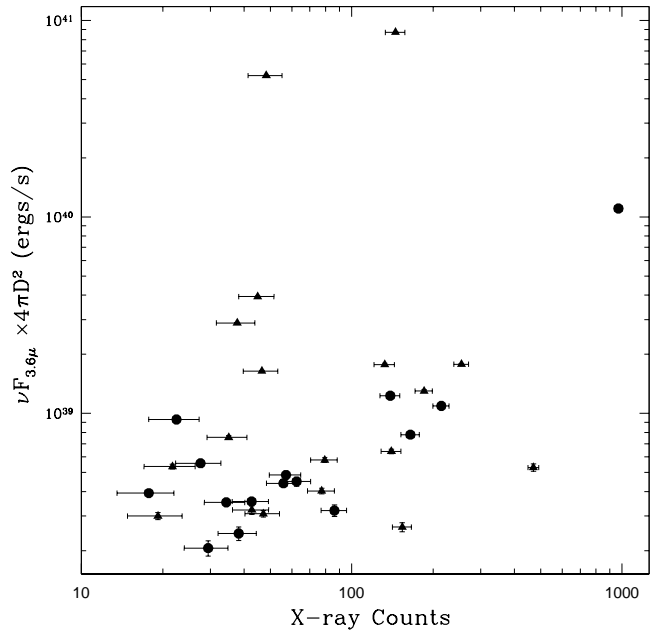}
\caption{IR(3.6$\mu$m) Luminosities v/s X-ray counts for the IR counterparts. The points with positive (log($F_{5.8}/F_{3.6}$)) and negative ratios are shown as circles and triangles respectively. The Spearman Rank correlation co-efficients are 0.49 (prob=0.06) and 0.12 (prob=0.65) respectively. For sources with X-ray counts greater than 50, the co-efficients become 0.76 (prob=0.03) and 0.17 (prob=0.67).}
\label{corrxc}
\end{figure}

\section{X-ray IR Correlations}

Figure \ref{corrxc} shows the IR 3.6$\mu$m luminosity, $L_{IR(3.6)} (\equiv \nu F_\nu 4 \pi D^2)$ v/s X-ray counts 
for the 33 X-ray point sources with IR point source counterparts. Overall there does not seem to be 
any correlation. However, for sources with AGN-like IR color i.e. $F_{5.8}/F_{3.6} > 1$ (circles) the Spearman rank correlation (significance level = 0.05) 
co-efficient is $0.49$ which corresponds to a probability of $0.06$. 
When 
we consider only brighter sources with X-ray counts $> 50$, 
the rank probability decreases to $0.03$. Sources that do not have 
AGN-like colours i.e. $F_{5.8}/F_{3.6} < 1$ (triangles) do not show any significant IR-X-ray correlation 
with a Spearman Rank correlation co-efficient of $0.12$ and probability of $0.65$.

\begin{figure}
\centering
\includegraphics[width=9cm,height=9cm]{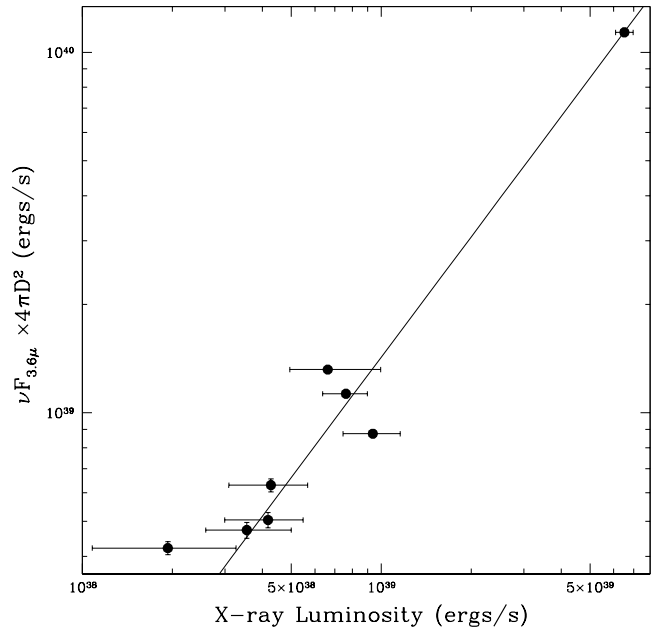}
\caption{ IR(3.6$\mu$) v/s unabsorbed X-ray luminosities for sources with positive (log($F_{5.8}/F_{3.6}$)) 
ratios. The correlation co-efficient is 0.91 (prob=0.002). Excluding the most luminous object at 3.6$\mu$m, the correlation co-efficient becomes 0.86 (prob=0.01).}
\label{corr36p}
\end{figure}

 Out of the 35 X-ray bright sources with spectra, 16 of them have mid-IR counterparts (see Figure \ref{image1}). 
 We compared the IR 3.6$\mu$m  with the unabsorbed X-ray 
luminosities for eight of the sources that have AGN-like colours in Figure \ref{corr36p}. The Spearman rank correlation is significantly better (assuming a significance level of 0.05) with a 
probability of $0.002$ in contrast to $0.007$ when X-ray counts are considered instead of the luminosity. 
Moreover, the correlation is given by
\begin{equation}
\frac{L_X}{10^{39} \hbox{ergs/s}} = (0.72 \pm 0.06) (\frac{L_{IR(3.6)}}{10^{39} \hbox{ergs/s}})^{0.90 \pm 0.04}
\end{equation}
with a reduced $\chi^2 = 0.86$. In contrast, fitting $L_{IR(3.6)}$ with X-ray counts, using the same 
parametric function gives a reduced $\chi^2 = 10.74$. 

\begin{figure}
\centering
\includegraphics[width=9cm,height=9cm,angle=0]{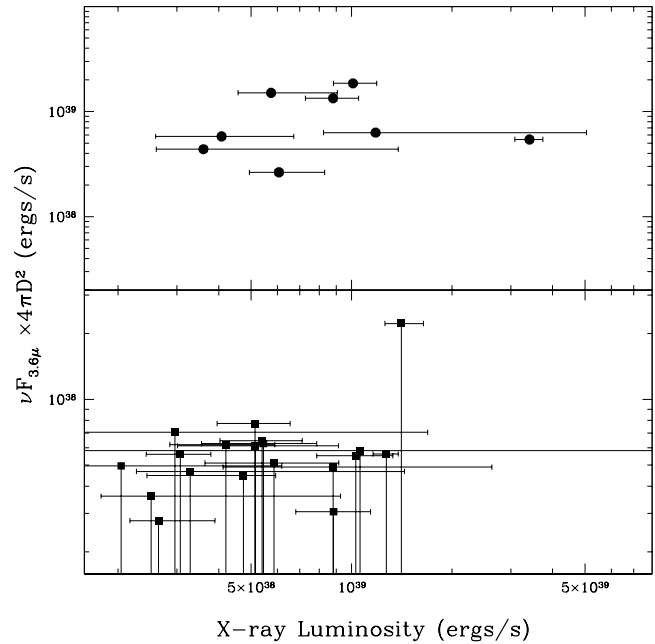}
\caption{IR(3.6$\mu$) v/s unabsorbed X-ray luminosities for sources with positive (log($F_{5.8}/F_{3.6}$)) ratios (above) and the 3-sigma upper limits for the 19 locations without IR counterparts (below). The correlation co-efficient is 0.29 (prob = 0.49).}
\label{corr36n}
\end{figure}

For eight sources which do not have AGN-like IR colours, the X-ray and the $3.6\mu$m IR luminosities do not 
correlate with each other as shown in the top panel of Figure \ref{corr36n}.

\begin{figure}
\centering
\includegraphics[width=9cm,height=9cm,angle=0]{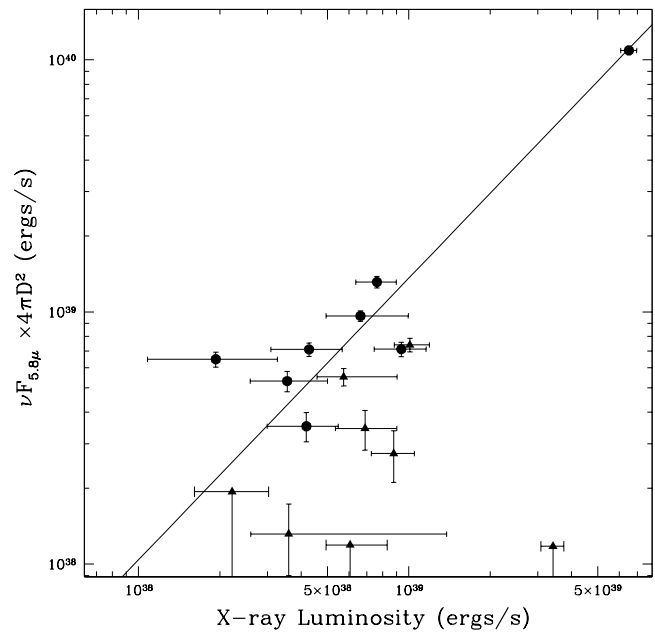}
\caption{IR(5.8$\mu$) v/s unabsorbed X-ray luminosities. The points with positive (log($F_{5.8}/F_{3.6}$)) and negative ratios are shown as circles and triangles respectively. The Rank correlation co-efficient for all the sources is 0.29 (prob=0.27). For the sources with positive and negative flux ratios, the co-efficients are 0.83 (prob=0.01) and 0.02 (prob=0.96) respectively. 3-sigma upper limits are plotted for the sources with fluxes less than this value 
.}
\label{corr2}
\end{figure}

Also for the AGN-like sources, the $5.8\mu$m luminosity 
is correlated with the X-ray luminosity (circles in Figure \ref{corr2}) and is given by
\begin{equation}
\frac{L_X}{10^{39} \hbox{ergs/s}} = (0.76 \pm 0.1) (\frac{L_{IR(5.8)}}{10^{39} \hbox{ergs/s}})^{0.89 \pm 0.06}
\end{equation}
which is consistent with the best fit parameters for $3.6\mu$m. However, the fit is not as
good here with a reduced $\chi^2 = 2.16$.
There is no such correlation seen between the  
the $5.8\mu$m and the X-ray luminosity for the eight sources without AGN-like colours (triangles in Figure \ref{corr2}).

For the remaining 19 bright X-ray sources there is no IR counterpart detected at the 3-sigma level. Figure 
\ref{corr36n}, shows the distribution of upper limits on the IR luminosities of these sources versus the 
X-ray luminosity. 

\begin{figure}
\centering
\includegraphics[width=9cm,height=9cm,angle=0]{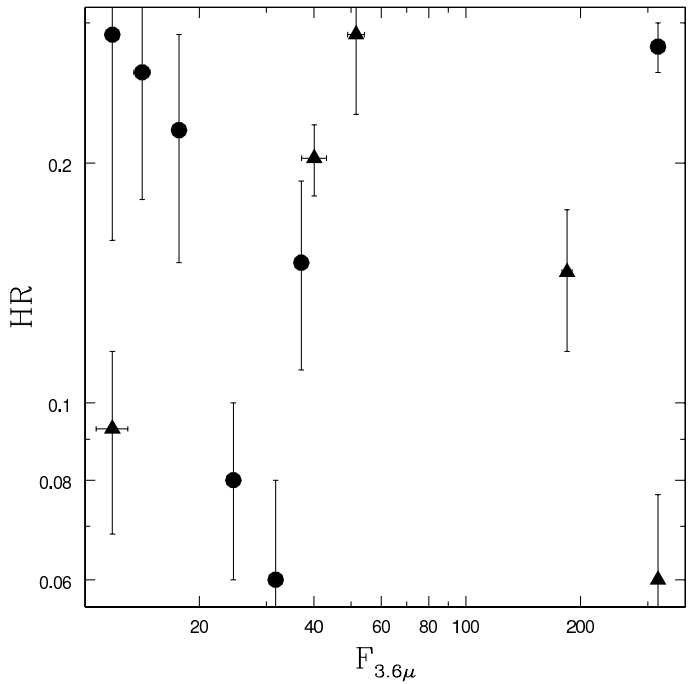}
\caption{IR(3.6$\mu$) fluxes v/s X-ray hardness ratios. The points with positive (log($F_{5.8}/F_{3.6}$)) and negative ratios are shown as circles and triangles respectively. The Rank correlation co-efficient for all the sources is -0.42 (prob=0.10) 
.}
\label{hr}
\end{figure}

 We have also looked for correlations of the IR flux with X-ray hardness ratios for the 16 detected IR sources. Figure \ref{hr} shows the plot of the hardness ratio against the 3.6$\mu$m fluxes. The Rank correlation co-efficient is -0.42 with a probability of 0.1 indicating a lack of correlation between the two.
\section{Optical counterparts identified from HST data}
The Optical analysis of NGC 1399 was carried out using the archival
images of HST with Advanced Camera for Survey (ACS). Among the archival
data sets, the observations with longer exposure time and multiple filters,
especially the wide band filters (F814W, F475W, F850LP and F606W) were
considered for the analysis. Images in different pointings were also taken
to have the maximum number of X-ray sources in the field of view. To
overcome the effect of the varying host galaxy background intensity on the
detection and photometry of the point sources, a smooth model of the galaxy was
subtracted from the observed image. The smooth model was generated by fitting
 elliptical isophotes to the brightness distribution of the galaxy using the
{\bf {\sc ellipse}} task in IRAF/STSDAS.
The other brighter objects near
the galaxy were masked during the fitting. The model image was subtracted
from the observed image (F814W) to get the residual image. The
object extraction was carried out on the residual image using SExtractor, to detect
and extract the sources for the 3-sigma detection threshold. When images
are available in multiple filters and different pointings, one of the optical
image is taken as the reference image (Obs ID : J9P305010) and the other images are
aligned to the reference one using {\bf {\sc geomap}} and {\bf {\sc geotran}} tasks in IRAF.
 We find that 26 of the {\it Chandra} X-ray sources are
within the field of view of available HST data sets and 21 sources have
an optical counterpart at least in one filter, within a positional offset of less than
one arc-second. This constant offset was applied to the {\it Chandra} source positions, thereby matching these shifted sources 
with the optical sources in the SExtractor catalogue. There are five
X-ray sources which do not have optical counterparts (at the 3-sigma level) within
one arc second of their shifted positions and are {\it optically-dark} X-ray
bright sources \citep{Jith11}.
\section{Optical Photometry}
Photometry of the identified optical counterparts of X-ray point sources were
performed on the drizzled image using the apphot package in IRAF. The drizzled images,
which are in units of $e^{-}/s/pixel$, are converted to
$e^{-}$ per pixel by multiplying by the total exposure times. An aperture
of radius 10 pixels was used to extract the flux in the task {\it APPHOT} and the Vega 
magnitude was computed with zero points taken from {\it HST ACS} data handbook. 
For the optically dark sources, we estimated the upper limit on their optical 
flux based on the 3-sigma detection threshold at that position.
\section{Optical Colours}
We have calculated the $g-z$ optical colours for 17 of the 35 X-ray bright sources. 
\citet{Paolillo2011} define the range ($1.3 < g-z < 2.5$) for which they consider 
an optical source to be a globular cluster. In our sample, 16 of the 17 sources with color
information may be globular clusters according to this criterion. Moreover, 7 of these 16 
sources are blue ($1.3 < g-z < 1.9$) while the rest 9 are red ($1.9 < g-z < 2.5$). 
\citet{Paolillo2011} reported that the majority of the X-ray sources reside in red globular 
clusters. However, it seems that for the bright X-ray sources considered here, a 
significant fraction of them are located in the blue clusters. Of the 8 X-ray sources 
with AGN-like IR colours (Table 1), 4 of them have $g-z$ values, and 3 of them are 
optically blue and one is red. Of the 8 sources which have non-AGN-like IR colours 
(Table 2), 5 of them have $g-z$ values. Here one is extremely blue ($g-z \sim 1$), 
two are blue ($g-z \sim 1.8$) and the other two are red ($g-z > 2$). Thus, these results indicate 
that although in general X-ray sources primarily reside in red globular clusters 
\citep{Paolillo2011}, the optical colours of sources with IR counterparts are primarily 
blue.  
This is particularly true for those sources which have AGN-like IR colours.

We have also compared the positions of the sources with the known Ultra Compact Dwarf 
Galaxies near NGC 1399, and found that none of these X-ray sources can be identified 
with them.

\section{Contamination by Background AGN}

It is possible that some of the X-ray sources considered in this work are distant background 
AGN since 40\% of the X-ray sources in elliptical galaxies are found to be background objects \citep{Swartz04,Paolillo2011}. This is particularly true for the 8 bright X-ray sources which have IR counterparts with 
AGN-like colours. If we assume a contamination of 40\% we find that 7 of the 16 sources could be background AGN and then the 
X-ray/IR correlation is just a manifestation of the known correlation for AGN \citep{asmus11}. 
Indeed, the fact that three of the four sources which have measured optical colours are blue, 
further indicates that they may be background AGN. If that is so, the IR flux ratio of the 
infra-red counterparts can be used as a good discriminator to distinguish background AGN from 
the X-ray sources within the galaxy. Following this argument, the X-ray bright sources that
belong to the host galaxy would have non-AGN IR colours. For example, one of the ULX (Source no. 
9, see Table 2) with an X-ray luminosity of $\sim 3.4 \times 10^{39}$ ergs/s has non-AGN like 
IR color, and its optical counterpart is consistent with being a red globular cluster. 
However, there are some indications that several of the sources showing AGN-like colours may not 
be background AGN. For example, the source no. 6 (Table 1) has IR flux ratio expected for AGN,
but has red optical colours. 
 We analysed the Spitzer IRAC image of NGC 4485/4490 and detected two of the sources studied 
by \citet{Vaz07} in both the IRAC wavelength channels (3.6$\mu$m and 5.8$\mu$m). For both  these sources
$F_{5.8}/F_{3.6} > 1$ indicating that they have AGN-like colours. Using emission line ratios, 
\citet{Vaz07} classified one of these sources as  AGN-like and the other as a supernova remnant. 
 The only decisive way to determine whether these objects are in the galaxy or background AGN is 
to obtain high quality spectra in IR/optical.
\section{Discussion}
We have analysed the Spitzer/IRAC images at 3.6$\mu$m and 5.8$\mu$m to identify mid-IR 
counterparts of the X-ray sources in NGC 1399. We considered the 35 bright X-ray sources for 
which the X-ray counts are high enough to perform reliable spectral modelling and estimate the
X-ray luminosities. We find that 16 of them are detected in the IR (see Figure \ref{image1}).
We have used the mid-IR colours of these IR sources to classify them into 
AGN-like and non-AGN-like sources. We find that for 8 sources that have AGN-like (F$_{5.8}/F_{3.6} > 1$) 
colours, the X-ray luminosities correlate well with the mid-IR luminosities as is the 
case for AGN. On the other hand the X-ray and mid-IR luminosities of the remaining sources do 
not show any correlation. 

Using the optical colours, we find that the AGN-like counterparts are mostly blue in contrast 
to the general X-ray sources that have preferentially red optical colours. While it appears 
that AGN-like IR flux ratios and blue optical colours can potentially be used to identify the 
background AGN, we cite the case of X-ray sources in NGC 4485/4490 where there are off-nuclear 
X-ray sources belonging to the host galaxy which exhibit AGN-like IR colours. The background 
AGN can only be reliably identified by spectral analysis.

In case a majority of the sources are AGN we re-affirm that mid-IR colours can be used as a 
good technique to differentiate AGN from other X-ray sources. However it is possible that the 
majority of them are sources in the galaxy itself. If this is the case, we extend the  
L$_{IR}$-L$_{X}$ correlation of \citet{asmus11} down to X-ray luminosities of $10^{39}$ ergs s$^{-1}$. 

 Another possibility is the contamination from nearby star-forming regions which could also show a correlation between IR and X-ray luminosities. However, this is highly unlikely in this case since there is no evidence for star-formation in NGC1399. Also for star-forming galaxies the IR fluxes are found to be much higher than their X-ray fluxes, while they are of the same order here.

In summary, our study of the mid-IR counterparts of X-ray sources shows that there are two 
categories of bright X-ray sources, one where the mid-IR luminosity correlates with the X-ray 
like AGN and the other where the IR and X-ray luminosities are uncorrelated. 
 This indicates  a possible difference in the 
environment of these two sources, i.e. one which has a dusty environment, other being dust deficient. 
The difference in the two classes of X-ray sources is further highlighted by optical photometry 
which suggests that bright X-ray sources that are AGN-like are bluer than typical X-ray sources 
which are red. This again perhaps indicates either a difference in environment or in the nature 
of the globular cluster that hosts the two classes.

However, as mentioned earlier these results need further investigations through IR spectroscopy to confirm what 
fraction of sources are background AGN or belong to the galaxy. The sources identified in this work 
are potential candidates for further spectroscopic follow-up studies. Moreover there is a 
need for a comparative study with other elliptical galaxies which will help us understand whether 
they host two classes of X-ray sources as seen in the case of NGC 1399. Also, a larger sample in
different galaxies could give a better understanding on the nature of the two classes of X-ray sources.

\section*{Acknowledgements}
We thank the referee for useful suggestions and comments which helped us to greatly improve the manuscript.
V.J., K.J., C.D.R., and B.R.S.B. thank the IUCAA visitors program and UGC Special assistance 
program. This work has been partially funded from the ISRO-RESPOND program. The authors thank 
Phil Charles for useful discussions. V.J. acknowledges financial support from the Council of Scientific and Industrial Research (CSIR) through SRF scheme.
P.S. would like to thank Calicut University for their support and hospitality.
This work is based [in part] on observations made with the Spitzer Space Telescope, which is operated by the Jet Propulsion Laboratory, California Institute of Technology under a contract with NASA.

\begin{landscape}
\begin{deluxetable}{lccccccccccccccr}
\tabletypesize{\small}
\label{tab1}
\tablecolumns{13}
\setlength{\tabcolsep}{2.3pt}
\tablenum{1}
\tablewidth{0pc}
\tablecaption{Properties of mid-IR counterparts with positive log($F_{5.8}/F_{3.6}$) ratios.}
\tablehead{
\colhead{No}&\colhead{R.A.}&\colhead{Dec.}&\colhead{m$_{814W}$}&\colhead{m$_{475W}$}&\colhead{m$_{850LP}$}&\colhead{m$_{606W}$}&\colhead{$L_{X}$}&\colhead{$\chi^{2}$/dof}&\colhead{3.6$\mu$m}&\colhead{5.8$\mu$m}&\colhead{$F_{5.8}/F_{3.6}$}&\colhead{$g-z$}\\
\colhead{(1)}&\colhead{(2)}&\colhead{(3)}&\colhead{(4)}&\colhead{(5)}&\colhead{(6)}&\colhead{(7)}&\colhead{(8)}&\colhead{(9)}&\colhead{(10)}&\colhead{(11)}&\colhead{(12)}&\colhead{(13)}}
\startdata
1&3 38 51.61& -35 26 43.59&-&-&-&19.89$\pm$0.001&$39.81^{+0.03}_{-0.02}$&55.78/58& $319.11\pm1.70$&$492.78\pm4.78$&$0.18$&-\\
2&3 38 25.29& -35 25 21.92&-&-&-&20.11$\pm$0.001&$38.88^{+0.07}_{-0.08}$&14.24/13& $31.73\pm0.71$&$59.42\pm3.10$&0.27&-\\
3&3 38 33.12& -35 27 31.29&20.698$\pm$0.005&21.806$\pm$0.007&20.499$\pm$0.008&21.418$\pm$0.004&$38.97^{+0.09}_{-0.10}$&20.47/11& $24.58\pm0.59$& $32.19\pm2.10$&$0.12$&1.307\\
4&3 38 45.35& -35 27 37.07&-&-&-&24.575$\pm$0.031&$38.82^{+0.18}_{-0.13}$&6.69/6& $37.03\pm0.46$&$43.62\pm2.09$&$0.07$&-\\
5&3 38 25.94& -35 27 41.77&25.729$\pm$0.00&26.773$\pm$0.00&25.437$\pm$0.00&27.383$\pm$0.00&$38.55^{+0.15}_{-0.14}$&2.37/3&$13.28\pm0.67$&$25.04\pm2.21$&$0.28$&-\\
6&3 38 32.34& -35 27 10.34&21.335$\pm$0.013&23.160$\pm$0.022&20.973$\pm$0.014&22.342$\pm$0.011&$38.63^{+0.12}_{-0.14}$&0.26/6& $17.70\pm0.73$&$32.11\pm1.95$&$0.26$&2.187\\
7&3 38 27.76& -35 27 50.2&22.983$\pm$0.045&24.194$\pm$0.049&22.818$\pm$0.055&23.926$\pm$0.033&$38.62^{+0.12}_{-0.15}$& 13.06/9&$ 14.17\pm0.70$& $15.92\pm2.12$&$0.05$&1.376\\
8&3 38 36.17& -35 26 25.11&21.429$\pm$0.008&22.528$\pm$0.010&20.872$\pm$0.009&21.760$\pm$0.004&$38.29^{+0.23}_{-0.25}$&4.40/4&$11.84\pm0.50$&$29.34\pm2.00$&$0.39$&1.656\\
\enddata
\tablecomments{(1) Source Number; (2) Right Ascension (J2000); (3) Declination (J2000); (4)-(7) Vega magnitudes in the HST F814W, F475W, F850LP and F606W filters; (8) unabsorbed X-ray luminosity (ergs/s); (9) $\chi^{2}$ value for X-ray model used to estimate the luminosity;(10), (11) IR flux in MJy for
the 3.6$\mu$m and 5.8$\mu$m bands; (12) mid-IR flux ratio; (13) optical $(g-z)$ color. The `-' sign denotes the
cases where sources are not in the FOV of the HST image.}
\end{deluxetable}
\end{landscape}

\begin{landscape}
\begin{deluxetable}{lcccccccccccccr}
\tabletypesize{\small}
\label{tab2}
\tablecolumns{13}
\setlength{\tabcolsep}{2.5pt}
\tablenum{2}
\tablewidth{0pc}
\tablecaption{Properties of mid-IR counterparts with negative log($F_{5.8}/F_{3.6}$) ratios.}
\tablehead{
\colhead{No}&\colhead{R.A.}&\colhead{Dec.}&\colhead{m$_{814W}$}&\colhead{m$_{475W}$}&\colhead{m$_{850LP}$}&\colhead{m$_{606W}$}&\colhead{$L_{X}$}&\colhead{$\chi^{2}$/dof}&\colhead{3.6$\mu$m}&\colhead{5.8$\mu$m}&\colhead{$F_{5.8}/F_{3.6}$}&\colhead{$g-z$}\\
\colhead{(1)}&\colhead{(2)}&\colhead{(3)}&\colhead{(4)}&\colhead{(5)}&\colhead{(6)}&\colhead{(7)}&\colhead{(8)}&\colhead{(9)}&\colhead{(10)}&\colhead{(11)}&\colhead{(12)}&\colhead{(13)}}
\startdata
9&3 38 32.60&-35 27 5.12&20.373$\pm$0.005&22.271$\pm$0.010&20.059$\pm$0.007&21.377$\pm$0.004&$39.53^{+0.04}_{-0.04}$&47.68/39&$15.25\pm0.68$&$<5.31$&$<-0.45$&2.21\\
10&3 38 48.71& -35 28  34.34&-&-&-&-&$39.00^{+0.07}_{-0.06}$&8.66/14& $52.11\pm0.49$& $33.46\pm2.07$&$-$0.19&-\\
11&3 38 20.05& -35 24 46.51&-&-&-&21.505$\pm$0.003&$38.94^{+0.08}_{-0.08}$&11.47/11&$37.65\pm0.67$&$12.41\pm2.87$&$-0.48$&-\\
12& 3 38 33.42&-35  23  2.62&-&-&-&-&$38.84^{+0.12}_{-0.11}$&4.63/5&$17.71\pm0.53$&$15.60\pm2.79$&$-0.06$&-\\
13&3 38 31.90& -35 26 49.04&21.061$\pm$0.011&21.992$\pm$0.009&20.994$\pm$0.014&21.757$\pm$0.007&$38.76^{+0.20}_{-0.10}$&11.90/9& $42.37\pm0.83$& $25.00\pm1.98$&$-0.23$&1.048\\
14&3 38 33.82& -35 25 56.43&19.373$\pm$0.002&20.984$\pm$0.003&19.171$\pm$0.003&20.216$\pm$0.002&$38.35^{+0.15}_{-0.14}$&3.51/3& $16.30\pm0.51$& $<8.76$ &$<-0.27$&1.813\\
15&3 38 38.80& -35 25  54.39&20.025$\pm$0.003&21.854$\pm$0.007&19.75$\pm$0.004&21.005$\pm$0.002&$38.56^{+0.58}_{-0.14}$& 3.71/2&$12.30\pm0.41$& $5.94\pm1.88$&$-0.32$&2.104\\
16& 3 38 36.83& -35 27 46.75&21.990$\pm$0.012&23.345$\pm$0.020&21.503$\pm$0.013&22.545$\pm$0.007&$38.78^{+0.14}_{-0.09}$&9.73/9& $7.41\pm0.42$& $<5.38$&$<-0.14$&1.842\\
\enddata
\tablecomments{(1) Source Number; (2) Right Ascension (J2000); (3) Declination (J2000); (4)-(7) Vega magnitudes in the HST F814W, F475W, F850LP and F606W filters; (8) unabsorbed X-ray luminosity (ergs/s); (9) $\chi^{2}$ value for X-ray model used to estimate the luminosity;(10), (11) IR flux in MJy for
the 3.6$\mu$m and 5.8$\mu$m bands; (12) mid-IR flux ratio; (13) optical $(g-z)$ color. The `-' sign denotes the
cases where sources are not in the FOV of the HST image.}
\end{deluxetable}
\end{landscape}

\begin{landscape}
\begin{deluxetable}{lcccccccccccr}
\tabletypesize{\small}
\label{tab3}
\tablecolumns{12}
\setlength{\tabcolsep}{2.5pt}
\tablenum{3}
\tablewidth{0pc}
\tablecaption{Properties of X-ray sources without mid-IR counterparts.}
\tablehead{
\colhead{No}&\colhead{R.A.}&\colhead{Dec.}&\colhead{m$_{814W}$}&\colhead{m$_{475W}$}&\colhead{m$_{850LP}$}&\colhead{m$_{606W}$}&\colhead{$L_{X}$}&\colhead{$\chi^{2}$/dof}&\colhead{3.6$\mu$m}&\colhead{5.8$\mu$m}&\colhead{$g-z$}\\
\colhead{(1)}&\colhead{(2)}&\colhead{(3)}&\colhead{(4)}&\colhead{(5)}&\colhead{(6)}&\colhead{(7)}&\colhead{(8)}&\colhead{(9)}&\colhead{(10)}&\colhead{(11)}&\colhead{(12)}}
\startdata
17. &3 38 31.82& -35 26 3.76&21.132$\pm$0.008&22.901$\pm$0.015&20.882$\pm$0.010&21.786$\pm$0.023& $39.10^{+0.04}_{-0.04}$&25.14/27&$<$1.69&$<$6.32&2.019\\
18. &3 38 27.64& -35 26 48.16&$<$25.121&$<$26.253&$<$24.339&$<$26.468&$39.29^{+0.09}_{-0.14}$&44.34/30&$<$6.70&$<$8.84&-\\
19.&3 38 29.68& -35 25  4.10 &-&-&-&-&$39.01^{+0.11}_{-0.12}$&8.22/9&$<$1.66&$<$8.80 &-\\
20.&3 38 21.91& -35 29  28.30&-&-&-&-&$38.95^{+0.11}_{-0.11}$&14.96/9& $<$0.92&$<$5.57&- \\
21.&3 38 31.73& -35 30  58.65&-&-&-&-&$38.42^{+0.17}_{-0.09}$&5.96/5& $<$0.84&$<$4.97&- \\
22.&3 38 26.50& -35 27 31.91&$<$25.402&$<$26.883&$<$25.343&$<$27.266&$38.71^{+0.11}_{-0.11}$&9.00/7&$<$2.33&$<$6.48&-\\
23.&3 38 32.80& -35 26 58.29&21.703$\pm$0.017&23.184$\pm$0.022&21.457$\pm$0.022&22.546$\pm$0.011&$38.74^{+0.16}_{-0.19}$&5.39/7& $<$1.89&$<$5.53&1.727\\
24.&3 38 31.28&-35 24 12.12&-&-&-&-&$38.95^{+0.48}_{-0.33}$&2.35/2&$<$1.47&$<$8.42&- \\
25.&3 38 42.42&-35 24 0.60&-&-&-&-&$38.73^{+0.12}_{-0.13}$&3.31/3&$<$1.95&$<$11.64&- \\
26.&3 38 15.45&-35 26 29.05&-&-&-&-&$38.52^{+0.50}_{-0.11}$&6.21/4&$<$1.41&$<$8.33&- \\
27.&3 38 32.35&-35 27 1.75&23.213$\pm$0.07&24.562$\pm$0.078&23.061$\pm$0.091&23.885$\pm$0.041& $38.47^{+0.76}_{-0.53}$&2.98/5& $<$2.13&$<$5.45&1.501\\
28.&3 38 33.80& -35 26 57.96&21.058$\pm$0.008&22.946$\pm$0.016&20.762$\pm$0.010&22.067$\pm$0.006&$39.03^{+2.95}_{-1.06}$& 0.46/2&$<$1.75&$<$5.62&2.184\\
29.&3 38 25.28& -35 27 52.97&20.519$\pm$0.004&22.37$\pm$0.009&20.255$\pm$0.006&21.52$\pm$0.004&$38.62^{+0.15}_{-0.17}$&2.08/3& $<$1.87&$<$6.41&2.115\\
30.&3 38 27.17& -35 26 0.91&20.434$\pm$0.004&22.319$\pm$0.010&20.137$\pm$0.006&21.479$\pm$0.004&$38.71^{+0.25}_{-0.23}$&5.29/3& $<$1.85&$<$9.17&2.182\\
31.&3 38 27.87&-35 25 26.85&$<$26.372&$<$26.803&$<$25.696&$<$27.361&$38.31^{+0.48}_{-0.32}$&2.53/4&$<$1.49&$<$6.79&- \\
32.&3 38 11.65&-35 26 48.41&-&-&-&-&$38.68^{+0.10}_{-0.29}$&0.85/4&$<$1.35&$<$7.34&- \\
33.&3 38 31.68& -35 26 0.22&20.814$\pm$0.006&22.789$\pm$0.014&20.464$\pm$0.008&21.836$\pm$0.005&$38.49^{+0.09}_{-0.10}$&6.76/10& $<$1.69&$<$6.30&2.325\\
34.&3 38 28.96& -35 26 2.0&23.456$\pm$0.064&24.836$\pm$0.091&23.023$\pm$0.066&24.343$\pm$0.042&$38.77^{+0.19}_{-0.21}$&5.49/8&$<$1.54&$<$5.42&1.813\\
35.&3      38   19.85 &     -35       28    45.48&-&-&-&-&$38.40^{+0.57}_{-0.15}$&4.82/3&$<$1.08&$<$5.87&- \\
\enddata
\tablecomments{(1) Source Number; (2) Right Ascension (J2000); (3) Declination (J2000); (4)-(7) Vega magnitudes
in the HST F814W, F475W, F850LP, and F606W filters; (8) X-ray luminosity (ergs/s); (9) $\chi^{2}$ value for X-ray model used to estimate the luminosity; (10), (11) IR flux in MJy for
the 3.6$\mu$m and 5.8$\mu$m bands; (12) optical $(g-z)$ color. The `-' sign denote the
cases where sources are not in the FOV of the HST image.}
\end{deluxetable}
\end{landscape}
\twocolumn
\bibliography{spitzer}
\end{document}